\documentclass[11pt]{article}
\usepackage{moriond,epsfig,bbold}

\newcommand{\be}{\begin{equation}}
\newcommand{\ee}{\end{equation}}
\newcommand{\bi}{\begin{itemize}}
\newcommand{\ei}{\end{itemize}}
\newcommand{\bc}{\begin{center}}
\newcommand{\ec}{\end{center}}
\newcommand{\bea}{\begin{eqnarray}}
\newcommand{\eea}{\end{eqnarray}}

\newcommand{\nn}{\nonumber}

\def    \raw           {\rightarrow}

\def    \part          {\partial}

\newcommand{\NK}{\widetilde{N}}
\newcommand{\mK}{\widetilde{m}}
\newcommand{\one}{\mathbb{1}}

\begin{document}
%\begin{flushright}
%IFT-UAM/CSIC-07-39} \\
%\end{flushright}
%\vspace{1cm}

\title{Symmetry Breaking in Six Dimensional Flux Compactification Scenarios}

\author{D. Hernandez, S. Rigolin$^*$  {\rm and} M. Salvatori} 
\address{Departamento de Fisica Teorica y IFT, Universidad Autonoma de
 Madrid, Madrid, Spain}

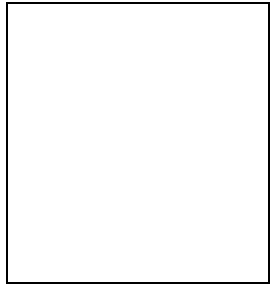
\maketitle\abstracts{
Motivated by the electroweak hierarchy problem, we consider theories with two extra dimensions 
in which the four-dimensional scalar fields are components of gauge boson in full space, 
namely the Gauge-Higgs unification framework. 
We briefly explain the basics features of "flux compactification", i.e. compactification 
in presence of a background (magnetic) flux. In particular we recall how chirality and 
symmetry breaking can be obtained in this context. More in details, we find and catalogue 
all possible degenerate zero-energy stable configurations in the case of trivial or non-trivial 
't Hooft flux, for a SU(N) gauge theory on a torus. 
We describe the residual symmetries of each vacua and the four-dimensional effective spectrum 
in terms of continuous and discrete parameters, respectively. }

\section{Introduction}
\label{sec:intro}
All available data agree in indicating that the mass of the Higgs boson is of the order of 
the Electroweak scale, $v\sim {\mathcal{O}}(200)$ GeV. Such a mass is unnaturally light if 
there is new physics beyond the Standard Model (SM) to which the Higgs boson is sensitive. 
In fact the Higgs mass parameter is not protected by any symmetry and thus gets corrections 
which are quadratically dependent on possible higher scales, like the unification scale 
$M_{GUT}$ or, ultimately, the Planck scale $M_{Pl}$.

Three different mechanisms have been devised in order to eliminate the quadratic sensitivity 
of the Higgs mass to the cutoff scale:
\begin{itemize}
 \item {\it Supersymmetry}: bosonic and fermionic contributions to the quadratic divergences 
       cancel each other in such a way that the Higgs mass remains affected only by a 
       logarithmic sensitivity to the cutoff scale;
 \item {\it Technicolor} and {\it Little Higgs}: the Higgs is a Goldstone boson of a global 
       custodial symmetry that it is only softly (spontaneously) broken;
 \item {\it Gauge-Higgs Unification}: the Higgs is a component of a higher dimensional gauge 
       multiplet. The lightness of its mass is protected by the gauge symmetry itself.
\end{itemize}
Independently of the precise nature assumed for the Higgs field, all these proposals require, 
in one way or another, the appearance of new physics at about the TeV scale. While the first 
two approaches are being intensely studied, in practice they tend to be afflicted by rather 
severe fine-tuning requirements when confronted with present experimental data. Here, instead, 
we concentrate on the last and less explored possibility: Gauge-Higgs unification \cite{Manton}.
The idea is that a single higher dimensional gauge field gives rise to all the four-dimensional 
($4D$) bosonic degrees of freedom: the gauge bosons, from the ordinary space-time components 
and the scalar bosons (and the Higgs fields among them) from the extra ones. The essential point 
concerning the solution of the hierarchy problem is that, although the higher dimensional gauge 
symmetry is globally broken by the compactification procedure, however it always remains locally 
unbroken. Any local (sensitive to the UV physics) mass term for the scalars is then forbidden by 
the gauge symmetry and the Higgs mass would then only have a non-local and UV finite origin.

This idea has been widely investigated in the context of five- and six-dimensional orbifold 
compactification \cite{recent}. From the field theory point of view, a different and less 
explored possibility is to recover the idea of Gauge-Higgs unification in the context of 
{\it flux compactification}: compactification of the extra space-like dimensions on a manifold 
in which there exist a (gauge) background with a non-trivial field strength, compatible with 
Scherk-Schwarz periodicity conditions \cite{SS}. We'll review in the following the basic 
idea of five-- and six-dimensional SS compactifications.

%%%%%%%%%%%%%%%%%%%%%%%%%%%%%%%%%%%%%%%%%%%%%%%%%%%%%%%%%%%%%%%%%%%%%%%%%%%%%%%%%%%%%%%
\section{Scherk-Schwarz mechanism in five- and six-dimensional compactifications}
\label{sec:intrinsic}
%%%%%%%%%%%%%%%%%%%%%%%%%%%%%%%%%%%%%%%%%%%%%%%%%%%%%%%%%%%%%%%%%%%%%%%%%%%%%%%%%%%%%%%

Let's consider a $U(N)$ gauge theory on a ($4+d$)-dimensional space-time \footnote{
Throughout the paper, with $x$ and $y$ we will denote the four-ordinary and $d$-extra 
coordinates, respectively. Latin upper case indices $M,N$ will run over all the 
extra-dimensional space, whereas greek and latin lower case indices $\mu,\nu$ and $a,b$ 
will run over the four ordinary and the extra-dimensions, respectively.} where the extra 
dimensions are compactified on an orthogonal $d$-dimensional torus $\mathcal{T}^d$. 
To completely define a field theory on a torus one has to specify the periodicity conditions: 
that is, to describe how the fields transform under the fundamental shifts $y \rightarrow 
y + l_a$, with $l_a$ being the lengths of the non-contractile cycles of the torus. Let's 
denote with $T_a$ the embeddings of these shifts in the fundamental representation of $U(N)$. 
If we want to preserve four-dimensional Poincar\'e invariance, the twists $T_a$ must depend 
only on the extra-dimensional coordinates. The most general periodicity conditions for 
the gauge field $A_M$ and for a generic field $\Phi$ in the fundamental representation of 
$U(N)$, read respectively:
\bea
\label{pcsgauge}
\mathbf{A_{M}} (x,y + l_a) &=& T_a (y) \left[ \,\mathbf{A_M} (x,y) \, + \, 
        \frac{i}{g} \partial_M  \right] T_a^\dagger(y) \, , \\
\Phi (x,y + l_a) &=& T_a (y) \, \Phi (x,y) \, .
\label{pcsfun}  
\eea
These equations are derived from the fact that while individual (gauge or matter) fields 
may not be single-valued on the torus, any physical scalar quantity, like the Lagrangian, 
must be. The periodicity conditions in Eqs.~(\ref{pcsgauge},\ref{pcsfun}) are usually referred 
as Scherk-Schwarz (SS) boundary conditions \cite{SS}. Let's describe in more details the 
five- and six- dimensional compactification procedure in presence of general SS boundary 
conditions.

%%%%%%%%%%%%%%%%%%%%%%%%%%%%%%%%%%%%%%%%%%%%%%%%%%%%%%%%%%%%%%%%%%%%%%%%%%%%%%%%%%%%%%%
\subsection{Scherk-Schwarz mechanism in five-dimensions}
%%%%%%%%%%%%%%%%%%%%%%%%%%%%%%%%%%%%%%%%%%%%%%%%%%%%%%%%%%%%%%%%%%%%%%%%%%%%%%%%%%%%%%%

In the case of a five-dimensional theory compactified on a circle $S^1$ one has to define a 
single twist matrix $T(y)$. No restrictions have to be imposed on $T$ except that it belongs 
to the $U(N)$ gauge group \footnote{The case of external automorphisms is not considered here. 
See for example \cite{Hebecker:2001jb}}. The $U(N)$ twist matrix can be, locally, decomposed 
as the product of an element $e^{i v(y)} \in U(1)$ and an element $\mathcal{V} (y) \in SU(N)$ 
as follows:
\be
T (y) \,\,=\,\,e^{i v(y)} \,\,\mathcal{V} (y)\,.
\ee
It is always possible to choose a gauge, called the {\em symmetric} gauge \cite{Hosotani}, 
in which the $SU(N)$ vacuum configurations are trivial and the twist matrix is constant 
$\mathcal{V}^{sym} = V$ and can be parametrized as:
\be
V = e^{2 \pi i (\alpha \cdot H)}  \quad  \quad , \quad  \quad 
\alpha \cdot H \equiv \sum_{j=1}^{N-1} \alpha^j H_j \, ,
\ee
where $H_{j}$ are the $(N-1)$ generators of the Cartan subalgebra of $SU(N)$ and $\alpha_j$ 
are $(N-1)$ real continuous parameters $0\leq \alpha_j < 1$. These parameters are non-integrable 
phases, which arise only in a topologically non-trivial space and cannot be gauged-away. 
When all the $\alpha_j$ are vanishing the periodicity conditions are trivial and consequently 
the initial symmetry is unbroken. If, instead, some of the $\alpha_j$ are non-vanishing, then 
symmetry breaking can occur. This mechanism is known as the {\em Scherk-Schwarz mechanism} 
\cite{SS}. 

In order to give an explicit expression for the gauge masses, one introduces the Cartan-Weyl 
basis for the $SU(N)$ generators. In addition to the Cartan subalgebra generators, $H_j$, 
one defines $N (N-1)$ non diagonal generators, $E_r$ such that the following commutation 
relations are satisfied:
\be
\left[ H_{j}, H_{k} \right] = 0 \quad , \quad 
%\,\,\,\,\,\,\,\forall j_1,j_2 = 1,...,N-1 ,,
\left[ H_{j}, E_r \right] = q^j_r E_r \,.
\label{CartanWeyl}
\ee 
In this basis, the twist $V$ acts in a diagonal way, that is 
\bea 
V H_j V^{\dagger} = H_j \quad & , & \quad V E_r V^{\dagger} = e^{2 \pi i \, 
                        (\alpha \cdot q_r) } \, E_r \,,
\eea
and the four-dimensional mass spectrum reads simply:
\be
m^2_{(k)} = \frac{ 4 \pi^2}{l^2} \left(n + \alpha \cdot q_k  
            \,\right)^2    \quad \quad , \quad \quad n \in \mathbb{Z} \,.
\label{SS5D}
\ee
For field components associated to a generator belonging to the Cartan subalgebra, $H_j$, 
one has $q_j = (0,...,0)$ and the spectrum reduce to the ordinary Kaluza-Klein (KK) one. 
For field components associated to the non-diagonal generators, $E_r$, one has, instead, 
$q_r \neq (0,...,0)$ and the mass spectrum is consequently shifted by a factor proportional 
to the non-integrable phases $\alpha^j$. When all the $\alpha^j\neq 0$, then only the gauge 
field components associated to the generators of the Cartan subalgebra are massless. Therefore, 
the symmetry breaking induced by the twists, $V$, does not lower the rank of $SU(N)$. 
The maximal symmetry breaking pattern that can be achieved for an $U(N)$ symmetry group 
is given by: 
\be
U(N) \sim U(1) \times SU(N) \rightarrow U(1) \times U(1)^{N-1} = U(1)^N.
\ee

Scherk-Schwarz symmetry breaking mechanism can be used to break both global (flavour 
symmetries, supersymmetry) or local symmetries. In the case of gauge symmetry breaking 
the SS phase, $\alpha$, can be interpret as the vev of the extra-dimensional component of 
the gauge fields, $\langle A_5 \rangle$. At classical level the scalar potential is flat 
and consequently the phases $\alpha^i$ are undetermined. Their values must be dynamically 
determined at the quantum level \cite{Luscher,Hosotani} minimizing the one-loop effective 
potential. If, at the minimum, any of the $\alpha_i$ is non-vanishing then the gauge 
symmetry is spontaneously broken. This dynamical and spontaneous symmetry breaking mechanism 
is conventionally known as the {\em Hosotani mechanism}. At the same time, the 
extra-dimensional component of the gauge field, $A_5$, is a scalar field that can be 
identified with the Higgs field and that acquires a finite mass term. The non-local nature 
of this symmetry breaking protects the theory from ultraviolet divergences and makes it a 
promising candidate mechanism to break the electroweak symmetry and to provide an Higgs 
field free from quadratic divergences. 

%%%%%%%%%%%%%%%%%%%%%%%%%%%%%%%%%%%%%%%%%%%%%%%%%%%%%%%%%%%%%%%%%%%%%%%%%%%%%%%%%%%%%%%
\subsection{Scherk-Schwarz mechanism in six-dimensions}
%%%%%%%%%%%%%%%%%%%%%%%%%%%%%%%%%%%%%%%%%%%%%%%%%%%%%%%%%%%%%%%%%%%%%%%%%%%%%%%%%%%%%%%

In the case of a six-dimensional theory compactified on a torus $T^2$, one can introduce 
a different twist, $T_a(y)$, along each of the two independent cycles. The twists cannot 
be chosen arbitrarily but they have to satisfy the following $U(N)$ 't Hooft consistency 
condition  \cite{'tHooft,Lebedev:1988wd}:
\be 
T_1 (y+l_2)\,T_2(y)\,\,= \,\, T_2 (y+l_1)\,T_1(y) \,.
\label{U(N)_cons_cond}
\ee
This condition is obtained imposing that (for any fields included in the theory) the value 
of the field at the final point $(y_1+l_1, y_2+l_2)$, starting from the initial point 
$(y_1,y_2)$ has to be independent on the followed paths. 

The $U(N)$ twist matrices can be, locally, decomposed as the product of an element $e^{i v_a(y)} 
\in U(1)$ and an element $\mathcal{V}_a (y) \in SU(N)$ as follows:
\be
T_a (y) \,\,=\,\,e^{i v_a(y)} \,\,\mathcal{V}_a (y)\,.
\label{T=ab_Omega}
\ee
Using this parametrization, the consistency conditions in Eq.~(\ref{U(N)_cons_cond}) can be 
splitted into the $SU(N)$ and $U(1)$ part, respectively:
\bea
\label{SU(N)_cons_cond}
\mathcal{V}_1 (y+l_2)\,\mathcal{V}_2(y) &=& e^{2 \pi i \frac{m}{N}} \,
\mathcal{V}_2 (y+l_1)\,\mathcal{V}_1(y) \\
\Delta_2 v_1(y) - \Delta_1 v_2(y)  &=&  2 \pi \frac{m}{N} \, , 
%\Delta_2 \,v_1(y) \,\,- \,\,\Delta_1 \,v_2(y) \,\,&=&  \,\,2 \pi \frac{m}{N} \, ,
%\,\hspace*{2em}\mathrm{modulo} \,\,\,\,N \,.
\label{U(1)_cons_cond}
\eea
with $\Delta_a v_b(y) = v_b(y+l_a) - v_b(y)$. The $SU(N)$ consistency condition,  Eq.~(\ref{SU(N)_cons_cond}), tells us that the twists, ${\mathcal V}_a$ must commute, 
on the fundamental plaquette, modulo a phase factor belonging to the center of $SU(N)$. 
The integer $m=0,1,..,(N-1)$ (modulo N) is a gauge invariant quantity called the 
non-abelian 't Hooft flux \cite{'tHooft}. Furthermore, Eq.~(\ref{U(1)_cons_cond}) tells 
us that it must coincide with the value of a quantized abelian magnetic flux living on 
the torus or, in other words, with the first Chern class of $U(N)$ on ${\cal T}^2$. 

%The phase in Eq.~(\ref{SU(N)_cons_cond}) is nothing else that the center of $SU(N)$. 
%The integer $m=0,1,..,N-1$ (modulo N) is a gauge invariant quantity called the 
%non-abelian 't Hooft flux \cite{'tHooft}. Furthermore, it coincides with the value of a 
%quantized abelian magnetic flux living on the torus or, in other words, with the first 
%Chern class of $U(N)$ on ${\cal T}^2$, Eq.~(\ref{U(1)_cons_cond}). 

It is well known, that the presence of a stable magnetic background, associated with 
the abelian subgroup $U(1) \in U(N)$ and living only on the two extra dimensions, can 
induce chirality \cite{Randjbar-Daemi:1982hi} in four-dimensions. A non-vanishing value 
of the 't Hooft flux $m$ is indeed necessary for having four-dimensional chiral matter 
fields. A general description of fermions and chirality in the context of 6D $U(N)$ 
theories compactified on a two-dimensional torus can be found in \cite{Faedo}.

From the other side, from Eqs.~(\ref{SU(N)_cons_cond},\ref{U(1)_cons_cond}) it appears 
evident that the presence of the quantized abelian magnetic flux deeply affects the 
non-abelian subgroup $SU(N) \in U(N)$, giving rise to a non-trivial 't Hooft non-abelian 
flux. While the symmetry breaking pattern for a $SU(N)$ theory in presence of trivial 
non-abelian 't Hooft flux ($m=0$) is well-known in the literature \cite{Luscher,Hosotani}, 
the field theory and phenomenological analysis of the non-trivial ($m \neq 0$) 't Hooft 
flux has been explored only recently, in \cite{Alfaro:2006is,Salvatori:2006pb}. Here, it 
has been shown that exists a gauge, denominated, as the five dimensional case, the {\it 
symmetric gauge}, in which the $SU(N)$ twists can always be chosen as constant, i.e.  $\mathcal{V}_a^{sym} = V_a$, with $V_a$ constant matrices satisfying the $SU(N)$ 't Hooft 
consistency conditions: 
\be 
V_1 \,V_2\,\,=\,\,e^{2 \pi i \frac{m}{N}}\,\,V_2\,V_1 \,.
\label{twist_algebra}
\ee 
In the symmetric gauge, the $SU(N)$ vacuum configurations are trivial and therefore the 
residual symmetries of each classical vacua are those associated to the $SU(N)$ generators 
which commute simultaneously with $V_1$ and $V_2$. The number of classical vacua and the 
pattern of symmetry breaking depend on the values of $m$ and they will be analyzed in the 
following section.

%%%%%%%%%%%%%%%%%%%%%%%%%%%%%%%%%%%%%%%%%%%%%%%%%%%%%%%%%%%%%%%%%%%%%%%%%%%%%%%%%%%%%%%%
\section{$ SU(N)$ Symmetry Breaking: trivial vs non-trivial 't Hooft flux}
\label{sec:eightfold}
%%%%%%%%%%%%%%%%%%%%%%%%%%%%%%%%%%%%%%%%%%%%%%%%%%%%%%%%%%%%%%%%%%%%%%%%%%%%%%%%%%%%%%%%

The main purpose this section is to find and classify all possible vacua and to describe 
the residual symmetries for an effective four-dimensional theory obtained from a $SU(N)$ 
gauge theory on a six-dimensional space-time where the two extra dimensions are compactified 
on a torus, for both the cases of trivial and non-trivial 't Hooft non-abelian flux. 

%%%%%%%%%%%%%%%%%%%%%%%%%%%%%%%%%%%%%%%%%%%%%%%%%%%%%%%%%%%%%%%%%%%%%%%%%%%%%%%%%%%%%%%%
\subsection{Trivial 't Hooft flux: $m=0$}
%%%%%%%%%%%%%%%%%%%%%%%%%%%%%%%%%%%%%%%%%%%%%%%%%%%%%%%%%%%%%%%%%%%%%%%%%%%%%%%%%%%%%%%%
In the $m=0$ case, Eq.~(\ref{twist_algebra}) tell us that the two $V_a$ matrices commute and 
consequently can be parametrized as:
\be
V_a= e^{2 \pi i (\alpha_a \cdot H)}  \quad  \quad , \quad  \quad 
\alpha_a \cdot H \equiv \sum_{j=1}^{N-1} \alpha_a^j H_j 
\label{V_a_m=0}
\ee
with $H_j$ the $(N-1)$ generators of the Cartan subalgebra of $SU(N)$. The periodicity 
conditions, and consequently the classical vacua, are now characterized by $2 (N-1)$ real 
continuous parameters, $0\leq \alpha_a^j < 1$. As in the five-dimensional case these 
parameters are non-integrable phases, which arise only in a topologically non-trivial 
space and cannot be gauged-away. When all the $\alpha_a^i$ are vanishing the initial 
symmetry is unbroken. At classical level $\alpha_a^i$ are undetermined. Their values 
must be dynamically determined at the quantum level where the rank-preserving Hosotani 
symmetry breaking mechanism can occur. 

The mass spectrum of the four-dimensional gauge and scalar components of the 6D gauge field 
follows straightfully the five-dimensional discussion. In the Cartan-Weyl basis Eq. 
\ref{CartanWeyl}, the twists $V_a$ act in a diagonal way, that is 
\bea 
V_a H_j V_a^{\dagger} = H_j \quad & , & \quad V_a E_r V_a^{\dagger} = e^{2 \pi i \, 
                        (\alpha_a \cdot q_r) } \, E_r \,,
\eea
and the four-dimensional mass spectrum for gauge/scalar fields reads:
\be
m^2_{(k)} = 4 \pi^2 \sum_{a=1}^2  \left(n_a + \alpha_a \cdot q_k  
            \,\right)^2 \frac{1}{l_a^2}   \quad \quad , \quad \quad n_a \in \mathbb{Z} \,.
\label{SS}
\ee
This is the same kind of spectrum seen previously in the five-dimensional case. For gauge 
(scalar) field components associated to a generator belonging to the Cartan subalgebra, 
$H_j$, the spectrum reduce to the ordinary Kaluza-Klein (KK) one. For gauge (scalar) field 
components associated to the non-diagonal generators, $E_r$, the mass spectrum is consequently 
shifted by a factor proportional to the non-integrable phases $\alpha_a^j$. Therefore, the 
symmetry breaking induced by the commuting twists, $V_a$, does not lower the rank of $SU(N)$. 

One can easily generalize these results to the $U(N)$ case adding an extra diagonal generator, 
$H_0 = \one_N/\sqrt{2N}$. Obviously $H_0$ commute with all the twists $V_a$ and consequently 
$A_M^0$ always remains unbroken. The maximal symmetry breaking pattern that can be achieved 
in the $m=0$ case, for an $U(N)$ gauge theory is given by: 
\be
U(N) \sim U(1) \times SU(N) \rightarrow U(1) \times U(1)^{N-1} = U(1)^N.
\ee
This symmetry breaking mechanism is exactly the same Hosotani mechanism one is used 
to in a five-dimensional framework.

%%%%%%%%%%%%%%%%%%%%%%%%%%%%%%%%%%%%%%%%%%%%%%%%%%%%%%%%%%%%%%%%%%%%%%%%%%%%%%%%%%%%%%%%
\subsection{Non-trivial 't Hooft flux: $m \neq 0$}
% %%%%%%%%%%%%%%%%%%%%%%%%%%%%%%%%%%%%%%%%%%%%%%%%%%%%%%%%%%%%%%%%%%%%%%%%%%%%%%%%%%%%%%
In the $m\neq0$ case, the twists $V_a$ don't commute between themselves and so necessarily 
they induce a rank-reducing symmetry breaking \cite{Salvatori:2006pb}. The most general 
solution of the consistency relation Eq.~(\ref{twist_algebra}) can be parametrized as follows: 
\be
V_1 = \omega_1 \,\,P^{s_1}\,Q^{t_1} \quad , \quad  V_2 = \omega_2 \,\,P^{s_2}\,Q^{t_2} \,.
\label{gaiarda}
\ee
$s_a, t_a$ are integers parameters taking values between $0,...,(N-1)$ (modulo $N$) and 
satisfying the following constraint: 
\be
s_1\,t_2 \,\,- \,\,s_2\,t_1\,\,=\,\,m/\mathcal{K} \equiv \mK\,.
\label{st_constraint}
\ee
$P$ and $Q$ are $SU(N)$ constant matrices given by
\bea
P \equiv P_{\NK} \otimes \one_{K} \quad & , &\quad Q \equiv Q_{\NK} \otimes \one_{K}
%P \equiv \left( 
%\begin{array}{cccc}
%P_{\NK} & 0 & ...& 0 \\
%0 & P_{\NK} & ...& 0 \\
%... & ... & ... & ...  \\
%0 & 0 & ... & P_{\NK}
%\end{array}
%\right)_{(\mathcal{K}\times \mathcal{K})} \quad & , &\quad 
%Q \equiv  \left( 
%\begin{array}{cccc}
%Q_{\NK} & 0 & ...& 0 \\
%0 & Q_{\NK} & ...& 0 \\
%... & ... & ... & ...  \\
%0 & 0 & ... & Q_{\NK}
%\end{array}
%\right)_{(\mathcal{K}\times \mathcal{K})} 
\label{PQ_paraculi}
\eea
where $\mathcal{K} \equiv \mathrm{g.c.d.} (m, N)$ and $\NK \equiv N/\mathcal{K}$.
$P_{\NK}$ and $Q_{\NK}$ are $\NK \times \NK$ matrices defined as 
\be
\left\{ \begin{array}{lcl}
\left(P_{\NK}\right)_{kj} & = & e^{i \pi \frac{\NK - 1}{\NK}} \,\, \delta_{k,j-1} \\
\left(Q_{\NK}\right)_{kj} & = & e^{- 2 \pi i \frac{(k-1)}{\NK}} \,\,e^{i \pi \frac{\NK-1}{\NK}}\,\, \delta_{kj}
\end{array} \right. \hspace*{2em} k,j\,=\,1,2,..., \NK,,
\label{PQ_def}
\ee
and satisfying the conditions 
\bea
P_{\NK}\,Q_{\NK} = e^{-2 \pi i \frac{1}{\NK}} Q_{\NK} P_{\NK} \quad & , & \quad 
\left(P_{\NK}\right)^{\NK} = \left(Q_{\NK}\right)^{\NK} = e^{\pi i (\NK-1)} \, .
\label{PQ_properties}
\eea
When ${\mathcal K}=1$, then $\NK =N$ and $P$, $Q$ reduce to the usual elementary twist 
matrices defined by 't Hooft in \cite{'tHooft}.

The matrices $\omega_a$ are constant elements of $SU(\mathcal{K}) \subset SU(N)$. They commute 
between themselves and with $P$ and $Q$. Therefore $\omega_a$ can be parametrized in terms 
of generators $H_j$ belonging to the Cartan subalgebra of $SU(\mathcal{K})$:
\be
\omega_a \,=\,e^{2 \pi i\, (\alpha_a \cdot H)} \quad \quad , \quad \quad 
\alpha_a \cdot H \equiv \sum_{\rho=1}^{\mathcal{K}-1}\,\alpha_a^\rho \,H_\rho
\label{omega_piccola}
\ee
Here $\alpha_a^\rho$ are $2 (\mathcal{K}-1)$ real continuous parameters, $0\le\alpha_a^\rho <1$.
As in the $m=0$ case, they are non-integrable phases and their values must be dynamically 
determined at the quantum level producing a dynamical and spontaneous symmetry breaking. 
%Indeed, we want to study, in presence of non-trivial 't Hooft flux, how the dynamics allows 
%to remove the degeneracy among the infinity of classical values of $\alpha_a^j$.

The $m\neq 0$ four-dimensional mass spectrum is easily obtained using the following basis \cite{Salvatori:2006pb} for the $SU(N)$ generators
\bea
\tau_{(\rho,\sigma)} (\Delta, k_\Delta) &=& \left\{ 
\begin{array}{lll}
\mathrm{if}\, \left\{\begin{array}{l}
\rho=\sigma \\
\Delta=k_\Delta=0
\end{array} \right.
 & \Rightarrow &\left(\sum_{i=1}^{\rho} \lambda^{\mathcal{K}}_{(i,i)} - 
   \rho \lambda^{\mathcal{K}}_{(\rho+1,\rho+1)}\right) \otimes \one_{\NK} \\ 
 & \\
\mathrm{else} & \Rightarrow & \lambda^{\mathcal{K}}_{(\rho,\sigma)} \otimes \tau^{\NK} (\Delta, k_\Delta) 
\end{array}
\right.
\label{base_fighetta}
\eea
where $\Delta,\,k_\Delta$ are integers assuming values between $0,\dots,(\NK-1)$ while the indices 
$\rho,\sigma$ take values between $1,\dots,\mathcal{K}$, excluding the case $\Delta=k_\Delta=0, 
\rho=\sigma$ in which $\rho$ takes values between $1,\dots,(\mathcal{K}-1)$. 
The matrices $\lambda^\mathcal{K}_{(\rho,\sigma)}$ and $\tau^{\NK}$ are $\mathcal{K} \times 
\mathcal{K}$ and $\NK \times \NK$ matrices, respectively, defined as:
\bea
\left(\lambda^{\mathcal{K}}_{(\rho,\sigma)}\right)_{ij} &=& \delta_{\rho i} \delta_{\sigma j} \nn \\
\tau^{\NK}(\Delta,k_{\Delta}) &=& \sum_{n=1}^{\NK} \, e^{2 \pi i  \frac{n}{\NK}\,k_\Delta}\,\lambda^{\NK}_{(n,n+\Delta)}  \,.
\label{gen_nodiag}
\eea
The definition of $\lambda^{\mathcal{\NK}}_{(n,n')}$ comes straightforwardly.

In this basis, the $SU(\mathcal{K})$ generators that commute with $P$ and $Q$ are simply given 
by $\tau_{\rho,\sigma} (0,0)$. In particular, the generators belonging to the Cartan subalgebra 
of $SU(\mathcal{K})$ are given by $H^{\rho}=\tau_{\rho,\rho}(0,0)$. The following commutation 
relations are satisfied:
\bea
\left[ \tau_{(\rho,\rho)}(0,0) , \tau_{(\rho,\rho)}(0,0) \right] &=& 0 \quad , \quad 
\left[ \tau_{(\rho,\rho)}(0,0), \tau_{(\sigma,\tau)}(\Delta,k_\Delta) \right] =
      q^{(\sigma,\tau)}_\rho \tau_{(\sigma,\tau)}(\Delta,k_\Delta) \,. \nn 
\eea
The action of the twists $V_a$ on this basis is given by
\bea
V_a \, \tau_{(\rho,\sigma)}(\Delta,k_\Delta) \,V_a^\dagger \,& = &\, 
  e^{\frac{2 \pi i}{\NK}(s_a \Delta + t_a k_\Delta)\,+ \, 2 \pi i \,
  ( \alpha_a \cdot q^{(\rho,\sigma)}) }\,\, \tau_{(\rho,\sigma)} (\Delta,k_\Delta)  \,,
\label{cho_K}
\eea
and the four-dimensional mass spectrum takes the following form:
\be
\label{m_spectrum_dani}
m^2_{(\rho,\sigma)}(\Delta,k_\Delta) = 4 \pi^2 \sum_{a=1}^2 \left(n_a + \frac{1}{\NK}\,(s_a\,\Delta\,\,+\,\,t_a\,k_\Delta) \,+\,
 \alpha_a \cdot q_{(\rho,\sigma)} \right)^2 \frac{1}{l_a^2}  \quad , 
%\quad n_a \in \mathbf{Z}.
\ee
with $n_a \in \mathbf{Z}$.
Therefore, beside the usual KK mass term, there are other two additional contributions. 
The first one, quantized in terms of $1/\NK$, is a consequence of the non-trivial commutation 
rule of Eq.~(\ref{PQ_properties}) between $P$ and $Q$ that induces the $SU(N) \rightarrow 
SU(\mathcal{K})$ symmetry breaking. Since $s_a,\,t_a$ cannot be simultaneously zero, 
the spectrum described by Eq.~(\ref{m_spectrum_dani}) always exhibits some (tree-level) 
degree of symmetry breaking. Given a set of $s_a, t_a$ and for all the $\alpha^\rho_a=0$ 
(that is $\omega_a=1$), only the gauge bosons components associated to $\tau_{(\rho,\sigma)}
(0,0)$, the generators of $SU(\mathcal{K})$, admit zero modes. This is an explicit breaking. 
The second contribution to the gauge mass comes from the $\omega_a$ and it depends on the 
continuous parameters $\alpha^\rho_a$. For $\mathcal{K}>1$ and all the non-integrable phases 
$\alpha_a^\rho \neq 0$, the only massless modes correspond to the gauge bosons 
associated to the Cartan subalgebra of $SU(\mathcal{K})$, i.e. $\tau_{(\rho,\rho)} (0,0)$. 
The symmetry breaking pattern induced by the $\omega_a$ produce a Hosotani symmetry breaking 
\cite{Faedo} that does not lower the rank of $SU(\mathcal{K})$. 

The maximal symmetry breaking pattern that can be achieved for an $U(N)$ gauge theory with matter 
fields in the fundamental is, in the $m\neq 0$ case, given by:
\be
U(N) \sim U(1) \times SU(N) \rightarrow U(1) \times U(1)^{\mathcal{K}-1} = U(1)^\mathcal{K}.
\ee
Obviously, when $\mathcal{K}=1$ the $SU(N)$ subgroup is completely broken, the only unbroken 
symmetry being the $U(1) \in U(N)$. This symmetry breaking pattern has no analogous in 
5-dimensional frameworks and it's peculiar of higher dimensional models where (topological) 
fluxes can appear. 

As a final comment on the spectrum, notice that in both the cases of trivial and non-trivial 
't Hooft flux, the classical effective four-dimensional spectrum depends on the gauge indices 
but it does not depend on the Lorentz ones. This implies that at the classical level the 
4D scalar fields $A_a$, arising from the extra-components of a six-dimensional gauge fields, 
are expected to be degenerate with the 4D gauge fields $A_\mu$ with the same gauge quantum 
numbers. This degeneracy is always removed at the quantum level \cite{Faedo}.

\section{Conclusions}
\label{sec:concl}

In this paper we have analyzed possible symmetry breaking mechanism in the context of 
Gauge-Higgs unification scenario. The introduction of general five-dimensional SS boundary 
conditions can drive a 4D gauge symmetry breaking through the dynamical mechanism, 
conventionally known as Hosotani mechanism. One-loop contributions to the scalar sector 
can shift the minimum of the effective potential and generate a non-vanishing vev for the 
Higgs field. This symmetry breaking is spontaneous and rank preserving. 

In six dimensions, SS boundary conditions have to satisfy a consistency condition. We 
discussed in details the $U(N)$ case where a novel ingredient appears: the non-abelian 
't Hooft flux. This flux is a topological quantity intimately connected with the U(1) 
(quantized) magnetic flux. In the case of trivial ($m=0$) 't Hooft flux the gauge 
symmetry breaking obtained thought SS boundary condition is the usual rank preserving 
Hosotani mechanism. In the case of non-trivial ($m\neq 0$) 't Hooft flux one can have, 
instead, two simultaneous symmetry breaking mechanism. A explicit, rank reducing, 
symmetry breaking associated to the non-commutativity of the twists leading to the 
$SU(N) \raw SU({\mathcal K})$ breaking. On top of that, for ${\mathcal K} > 1$, the 
residual symmetry group can be further reduced through a spontaneous, rank preserving, 
Hosotani mechanism. 

The simultaneous presence of rank preserving and rank reducing symmetry breaking mechanism 
makes the non-trivial 't Hooft flux case particularly interesting from a model building 
point of view.

%%%%%%%%%%%%%%%%%%%%%%%%%%%%%%%%%%%%%%%%%%%%%%%%%%%%%%%%%%%
\vspace{-0.25cm}
\section*{Acknowledgments}
We acknowledge E. Alvarez, A. Faedo, B. Gavela and D. Hernandez for useful discussions 
The work of S. Rigolin and M. Salvatori was partially supported by CICYT through the 
project FPA2003-04597 and by CAM through the project HEPHACOS, P-ESP-00346. M. Salvatori 
also acknowledges MECD for financial support through FPU fellowship AP2003-1540.

\end{document}